\documentclass[12pt]{article}

\usepackage{amssymb}

\begin{document}

\begin{titlepage}

\begin{flushright}
IUHET-488\\
hep-th/0512090
\end{flushright}
\vskip 2.5cm

\begin{center}
{\Large \bf Radiatively Induced Lorentz-Violating\\
Photon Masses}
\end{center}

\vspace{1ex}

\begin{center}
{\large B. Altschul\footnote{{\tt baltschu@indiana.edu}}}

\vspace{5mm}
{\sl Department of Physics} \\
{\sl Indiana University} \\
{\sl Bloomington, IN 47405 USA} \\

\end{center}

\vspace{2.5ex}

\medskip

\centerline {\bf Abstract}

\bigskip

We examine the radiative corrections to an extension of the standard model
containing a Lorentz-violating axial vector parameter.
At second order in this parameter, the photon self-energy is known to contain
terms that violate gauge invariance. Previously, this has been treated as a
pathology, but it is also possible to take the gauge noninvariant terms at face
value. These terms then make Lorentz-violating contributions to the photon mass,
and directly measured limits on the photon mass can be used to set bounds on
the Lorentz violation at better than the $10^{-22}$ GeV level.

\bigskip

\end{titlepage}

\newpage

There has recently been a great deal of interest in the possibility of there
existing small Lorentz- and CPT-violating corrections to the standard
model. Any observed violations of these fundamental symmetries would be
important clues
about the nature of Planck scale
physics. Within the context of effective field theory, a general
Lorentz-violating extension of the standard model has been
developed~\cite{ref-kost1,ref-kost2}.
However, the general standard model extension (SME) is extremely complicated,
and usually only restricted subsets of the SME (such as the minimal SME,
which is gauge invariant and superficially renormalizable) are
considered in the literature.
The minimal SME provides a framework within which to analyze the
results of experiments testing Lorentz violation. To date, such
experimental tests have included studies of matter-antimatter asymmetries for
trapped charged particles~\cite{ref-bluhm1,ref-bluhm2,ref-gabirelse,
ref-dehmelt1} and bound state systems~\cite{ref-bluhm3,ref-phillips},
determinations of muon properties~\cite{ref-kost8,ref-hughes}, analyses of
the behavior of spin-polarized matter~\cite{ref-kost9,ref-heckel},
frequency standard comparisons~\cite{ref-berglund,ref-kost6,ref-bear},
measurements of neutral meson oscillations~\cite{ref-kost7,ref-hsiung,ref-abe},
polarization measurements on the light from distant galaxies~\cite{ref-carroll1,
ref-carroll2,ref-kost11},
and others.

Radiative corrections to the SME are also a very interesting
subject~\cite{ref-kost4}. A great deal of work in this area has concerned the
corrections to the electromagnetic Chern-Simons term, with Lagrange density
${\cal L}_{CS}=\frac{1}{2}(k_{CS})_{\mu}\epsilon^{\mu\alpha\beta\gamma}
F_{\alpha\beta}A_{\gamma}$~\cite{ref-carroll1,ref-jackiw2,ref-schonfeld}, because
${\cal L}_{CS}$ is not gauge invariant (although the related action is).
The relevant minimal SME Lagrange density
for studying these questions is~\cite{ref-kost4}
\begin{equation}
\label{eq-L}
{\cal L}=-\frac{1}{4}F^{\mu\nu}F_{\mu\nu}+\bar{\psi}(i\!\!\not\!\partial-m-
e\!\!\not\!\!A\,-\!\not\!b\gamma_{5})\psi.
\end{equation}
This theory has the potential to induce a finite
radiatively-generated Chern-Simons
term, with $\Delta k_{CS}$ proportional to $b$. However, the coefficient of
proportionality depends upon the
regularization~\cite{ref-jackiw1,ref-victoria1,ref-victoria2}.  Different
regulators lead to different values of $\Delta k_{CS}$, and
through a suitable choice, any coefficient of proportionality between the two
may
be found.  This ambiguity has been extensively studied, and several potentially
interesting values of $\Delta k_{CS}$ have been identified~\cite{ref-victoria1,
ref-coleman,ref-chung1,ref-chung2,ref-chen,ref-chung3,ref-andrianov}. The
ambiguity can be related to the choice of momentum routings in the two triangle
diagrams that contribute to $\Delta k_{CS}$ at leading order. However, the
Chern-Simons term is not the only radiative correction in this theory with
quite peculiar gauge invariance properties.

The induced Chern-Simons term is a part of the full, Lorentz-violating photon
self-energy $\Pi^{\mu\nu}(p)$.
However, evaluation of this quantity is tricky.
The exact fermion propagator,
\begin{equation}
S(l)=\frac{i}{\!\not l-m\,-\!\not\!b\gamma_{5}},
\end{equation}
may be rationalized to obtain~\cite{ref-victoria1,ref-chung1}
\begin{equation}
\label{eq-propagator}
S(l)=i\frac{(\!\not l+m\,-\!\not\!b\gamma_{5})(l^{2}-m^{2}-b^{2}+[\!\not l,
\!\not\!b\,]\gamma_{5})}{(l^{2}-m^{2}-b^{2})^{2}+4[l^{2}b^{2}-(l\cdot b)^{2}]}.
\end{equation}
At $l=0$, the denominator of the rationalized propagator becomes
$(m^{2}+b^{2})^{2}$. The square root $|m^{2}+b^{2}|$ of this expression arises in
the evaluation of fermion-antifermion loop diagrams, and the absolute value leads
to
behavior that is nonanalytic in $b$. There may be other thresholds for
nonanalytic behavior in the photon self-energy as well.

Radiative corrections other than those to the Chern-Simons term are also
interesting.
We have previously~\cite{ref-altschul1,ref-altschul2} calculated the
${\cal O}(e^{2}b^{2})$ contribution to the photon self-energy in two different
regimes---that in which the fermions are massless, $m=0$; and that in which the
mass $m$ is much larger than the scale of the Lorentz violation, or $m^{2}\gg
|b^{2}|$. In the latter regime, we also restricted our attention to the
zero-momentum case, $p=0$. In either of these situations, a power series expansion
in $b$
is justified, and so it is not too surprising that the same result is found in
either case. The ${\cal O}(e^{2}b^{2})$ part of $\Pi^{\mu\nu}$ is
\begin{equation}
\label{eq-Pib2}
\Pi^{\mu\nu}_{b^{2}}= -\frac{e^{2}}{24\pi^{2}}\left(2b^{\mu}b^{\nu}+
g^{\mu\nu}b^{2}\right).
\end{equation}
This violates the Ward identity that enforces
transversality---$p_{\nu}\Pi^{\mu\nu}(p)=0$; however, this result is unambiguous,
as there are no momentum routing ambiguities.

Previously, we have treated this result for $\Pi^{\mu\nu}_{b^{2}}$ as a
pathology---as an indication that this method of regulating the theory was
inadeqaute. The failure of the Ward identity corresponds to a breakdown of gauge
invariance, so this is certainly a reasonable point of view to take.  However, it
is also possible to consider a different formulation of the theory, one in which
the radiative breakdown of gauge symmetry is a real effect.  Gauge invariance
can only be restored to the theory by introducing a nontrivial regulator for the
${\cal O}(e^{2}b^{2})$
terms in the Feynman diagram expansion. However, while these
terms have naive logarithmic divergences, they are unavoidably finite, so the
regulator is not strictly necessary.  That is, the role of the regulator is no
longer to render an infinite result finite, but merely to preserve gauge
invariance.  We shall therefore dispense with the regulator and take the
Ward-identity-violating terms at face value. If this produces a well-defined
theory, we may relate our results to experimental tests of special relativity.
However, any bounds we find on $b$ will not be as general as those arising from
direct measurements of $b$, because we have considered only one possible version
of the theory---one in which the radiative corrections are defined in a particular
way.

It is plausible, although unproven, that the result (\ref{eq-Pib2})
may actually hold for all
values of $b$ and $m$, so long as $p=0$;
this would occur if the nonanalyticities mentioned above
turn out not to be a problem at this order in $b$. However, we shall restrict
our attention to the situations in which we know that expanding $\Pi^{\mu\nu}$
as a power series in $b$ is justified. In any case, we expect $m^{2}\gg|b^{2}|$
to represent the physical regime for all charged particles.

In the $m=0$ case, (\ref{eq-Pib2}) holds for all values of $p$. However, in the
massive case, we should expect there to be additional momentum-dependent terms
that arise for $p\neq0$. However, terms involving positive powers of $p$ are
power-counting finite. These finite terms must automatically obey the Ward
identity, unlike the expression (\ref{eq-Pib2}); they possess the Lorentz
structure~\cite{ref-bonneau}
\begin{equation}
P^{\mu\nu}=g^{\mu\nu}p^{2}b^{2}-p^{\mu}p^{\nu}b^{2}-g^{\mu\nu}(p\cdot b)^{2}
-p^{2}b^{\mu}b^{\nu}+(p\cdot b)(p^{\mu}b^{\nu}+p^{\nu}b^{\mu}).
\end{equation}
Just like the usual transverse
Lorentz structure $g^{\mu\nu}p^{2}-p^{\mu}p^{\nu}$
that appears in the photon self-energy,
$P^{\mu\nu}$ is proportional to a projector:
\begin{equation}
\label{eq-proj}
P^{\mu\rho}P_{\rho}\,^{\nu}=[p^{2}b^{2}-(p\cdot b)^{2}]P^{\mu\nu}.
\end{equation}
Because of (\ref{eq-proj}) and the fact that the momentum-dependent terms
satisfy the Ward identity, we shall not consider these terms
here, although they still might lead to interesting effects. We shall
merely note that these terms will not possess poles at $p^{2}=0$. When the full,
momentum-dependent self-energy is evaluated
at ${\cal O}(e^{2}b^{2})$, with the introduction of a
Feynman parameter $x$, the momentum invariant that appears in the denominator of
the expression is the usual $m^{2}-x(1-x)p^{2}$. There cannot therefore be a pole
at $p^{2}=0$ unless $m=0$. However, the full momentum dependence is known in
exactly that case, and there is no pole present for $m=0$
either. So, since they lack any
$p^{2}=0$ poles, the momentum-dependent terms will not shift the
mass parameter that appears in the photon propagator.

There will be a mass generated by (\ref{eq-Pib2}), however. We shall demonstrate
this by resumming the geometric series of one-particle irreducible diagrams
\begin{equation}
D^{\mu\nu}_{b^{2}}(p)=\frac{-ig^{\mu\nu}}{p^{2}}+\frac{-ig^{\mu}\,_{\rho}}{p^{2}}
i\Pi^{\rho\sigma}_{b^{2}}\frac{-ig_{\sigma}\,^{\nu}}{p^{2}}+
\frac{-ig^{\mu}\,_{\rho}}{p^{2}}i\Pi^{\rho\sigma}_{b^{2}}\frac{-ig_{\sigma
\kappa}}
{p^{2}}i\Pi^{\kappa\tau}_{b^{2}}\frac{-ig_{\tau}\,^{\nu}}{p^{2}}+\cdots.
\end{equation}
The bare photon propagator has been left in the Feynman gauge.
Determining $D^{\mu\nu}_{b^{2}}$
involves the evaluation of the ``$n$-th power'' of $\Pi^{\mu\nu}_{b^{2}}$:
\begin{equation}
\left(\Pi_{b^{2}}^{(n)}\right)^{\mu\nu}=\left(\Pi_{b^{2}}\right)^{\mu}
\,_{\alpha_{1}}\left(\Pi_{b^{2}}\right)^{\alpha_{1}}\,_{\alpha_{2}}\cdots
\left(\Pi_{b^{2}}\right)^{\alpha_{n-1}\nu},
\end{equation}
where there are $n$ terms on the right-hand side. This expression is
straightforward to evaluate, and we find
\begin{equation}
\left(\Pi_{b^{2}}^{(n)}\right)^{\mu\nu}=\left(-\frac{e^{2}b^{2}}{24\pi^{2}}\right)
^{n}\left[(3^n-1)\frac{b^{\mu}b^{\nu}}{b^{2}}+g^{\mu\nu}\right].
\end{equation}
Inserting this into the expansion of $D_{b^{2}}^{\mu\nu}$ then gives
\begin{eqnarray}
D^{\mu\nu}_{b^{2}}(p) & = & \sum_{n=0}^{\infty}\frac{-i}{p^{2}}\frac{
\left(\Pi_{b^{2}}^{(n)}\right)^{\mu\nu}}{(p^{2})^{n}} \\
& = & \frac{-i}{p^{2}}\sum_{n=0}^{\infty}\left(-\frac{e^{2}b^{2}}{24\pi^{2}}
\frac{1}{p^{2}}\right)^{n}\left[3^{n}\frac{b^{\mu}b^{\nu}}{b^{2}}+
\left(g^{\mu\nu}-\frac{b^{\mu}b^{\nu}}{b^{2}}\right)\right].
\end{eqnarray}
The two Lorentz structures, which project out vectors parallel
and perpendicular to
$b$, generate separate geometric series. When summed individually, they give
\begin{equation}
D^{\mu\nu}_{b^{2}}(p)=-i\frac{b^{\mu}b^{\nu}/b^{2}}{p^{2}+\frac{e^{2}b^{2}}
{8\pi^{2}}}
-i\frac{g^{\mu\nu}-b^{\mu}b^{\nu}/b^{2}}{p^{2}+\frac{e^{2}b^{2}}
{24\pi^{2}}}.
\end{equation}
So this does indeed look like a massive theory, although there is not a single
mass, but rather two. For the component of $A$ in the same direction as $b$, the
pole in the propagator is shifted to $p^{2}=m^{2}_{\parallel}=-\frac{e^{2}b^{2}}{8
\pi^{2}}$, while the mass
sqaured parameter for the other components of the gauge field is
$m^{2}_{\perp}=-\frac{e^{2}b^{2}}{24\pi^{2}}$.

These are exactly the two masses that appear in the effective Lagrangian.
Including
only these gauge-noninvariant radiative corrections, the effective Lagrange
density for the purely electromagnetic sector becomes
\begin{eqnarray}
{\cal L}_{b^{2}} & = & -\frac{1}{4}F^{\mu\nu}F_{\mu\nu}-\frac{e^{2}}{24\pi^{2}}
(A\cdot b)^{2}-\frac{e^{2}b^{2}}{48\pi^{2}}A^{2} \\
\label{eq-massterms}
& = &
-\frac{1}{4}F^{\mu\nu}F_{\mu\nu}+
\frac{1}{2}\left[-\frac{e^{2}b^{2}}{8\pi^{2}}\frac{b^{\mu}
b^{\nu}}{b^{2}}-\frac{e^{2}b^{2}}{24\pi^{2}}\left(g^{\mu\nu}-\frac{b^{\mu}b^{\nu}}
{b^{2}}\right)\right]A_{\mu}A_{\nu}.
\end{eqnarray}
These kinds of Lorentz-violating photon mass operators do not appear in the
minimal SME, because they are not gauge invariant. Yet despite the gauge
invariance of the underlying bare action,
terms of the form $M^{\mu\nu}A_{\mu}A_{\nu}$ may yet appear in the effective
Lagrangian at one-loop and higher orders.

What we have found looks
like a Lorentz-violating (but CPT-preserving) variation of the Proca
Lagrangian for massive photons,
\begin{equation}
{\cal L}_{m_{\gamma}}=-\frac{1}{4}F^{\mu\nu}F_{\mu\nu}+\frac{1}{2}m_{\gamma}^{2}
A^{2}.
\end{equation}
A Lorentz-violating mass term more similar to (\ref{eq-massterms})
has also been considered in the
literature~\cite{ref-gabadadze,ref-dvali}---
\begin{equation}
{\cal L}'_{m_{\gamma}}=-\frac{1}{4}F^{\mu\nu}F_{\mu\nu}-\frac{1}{2}m_{\gamma}^{2}
\vec{A}^{2},
\end{equation}
where $\vec{A}^{2}=A_{j}A_{j}$ is the square of the three-vector potential.
(Somewhat more general Lorentz-violating photon mass terms are considered
in the appendix.)
In ${\cal L}'_{m_{\gamma}}$, there is one component of the field with a different
mass parameter. The electrostatic potential $A^{0}$ is not coupled to
$m_{\gamma}$, and so the Coulomb field is not screened. Bounds of the photon mass
are frequently found by observing a lack of screening in the fields of static
sources, and so unless these experiments are very carefully formulated, they may
really only be constraining the smallest mass parameter in the
photon Lagrangian.

In the model we have considered,
a spacelike $b$ is favored, since it generates a positive
definite mass matrix for the vector degrees of freedom. A timelike $b$ generates
tachyonic mass parameters, which render the theory unstable; in the appendix,
we show that a timelike $b$ leads to solutions with fields that grow exponentially
in time. These instabilities could be prevented by including
higher-order gauge-noninvariant terms, such as $(A^{2})^{2}$, because those terms
can make large values of the field energetically impossible.
Since this theory does not possess a fully gauge-invariant
effective action, such terms are not strictly forbidden. However, neither is it
clear that there is any mechanism by which they could arise.
For example, any $b$-dependent contribution to the ${\cal O}(e^{4})$
four-photon amplitude is finite by power-counting and so will not break gauge
symmetry. So without further modifications, the theory will only be stable if
$b$ is spacelike or lightlike (in which case, the mass parameters obviously
vanish). However, we should point out that even if
$b$ is timelike, so that there is an instability, which an ${\cal O}(A^{4})$
term then stabilizes, the size of the physical mass parameters will still be of
order $\frac{e^{2}b^{2}}{24\pi^{2}}$. (For comparison, a
timelike Chern-Simons coefficient also leads to instabilities; however, for the
fermionic $b$ itself, a timelike value is actually preferred.)

For a spacelike $b$, there are indeed physical states
with two distinct masses. By
mass, we mean precisely the energy eigenvalue at zero momentum, considered in
a frame in which $b$ is purely spacelike---$b^{\mu}=\left(0,\vec{b}\right)$.
In the absence of sources, and with all spatial derivatives in the equations of
motion vanishing, we have
\begin{equation}
\label{eq-zerop}
-\partial_{0}^{2}A_{j}=m_{\perp}^{2}A_{j}+2m_{\perp}^{2}\left(\vec{A}\cdot
\hat{b}\right)\hat{b}_{j},
\end{equation}
where $\hat{b}$ is the unit vector $\vec{b}/\left|\vec{b}\right|$. The two
components of $\vec{A}$ normal to $\vec{b}$ have mass $m_{\perp}$, as expected,
and the parallel component has the larger mass $m_{\parallel}$.
Supplementing (\ref{eq-zerop}) is a gauge condition derived from current
conservation, $\partial_{\mu}j^{\mu}=0$. The general condition is
\begin{equation}
\partial_{\mu}A^{\mu}+2\left(\partial_{\mu}A_{\nu}\right)\frac{b^{\mu}b^{\nu}}
{b^{2}}=0.
\end{equation}
In the zero-momentum case we are considering, this becomes $\partial_{0}A^{0}=0$,
and so it constrains $A^{0}$ to be constant in time. Dispersion relations with
more general mass terms and nonzero momenta are considered in the appendix.

Within the framework of our model, bounds on the photon mass may be translated
into bounds on $b$.
One aspect of this model that is particularly interesting is that it gives us
experimental access to the $b$ coefficients for all the charged particles in the
theory. Of course, the same thing is true in principle with the radiatively
induced Chern-Simons term. However, in that case there are additional
complications. Since the Chern-Simons term does not break the gauge invariance of
the action, there can be a classical contribution to $k_{CS}$ in addition to the
quantum contribution $\Delta k_{CS}$. It is impossible to disentangle the
classical and radiative parts completely, and this is further complicated by
the ambiguity in the radiatively induced part.
Moreover, there is good reason to
expect that there may be a cancelation among the different species. This
can occur if the $b$
terms ultimately arise from the vacuum expectation value of a quantized axial
vector field, $A_{5}^{\mu}$~\cite{ref-kost2,ref-chung2}.
Then for each species of fermions,
$b^{\mu}=g\langle A_{5}^{\mu}\rangle$, where $g$ is the coupling of that
species to $A_{5}^{\mu}$. If there is no anomaly associated with the
axial vector field, then the anomaly cancelation condition, when multiplied
by $\langle A_{5}^{\mu}\rangle$, gives
\begin{equation}
\sum_{f}q_{f}^{2}b_{f}^{\mu}=0.
\end{equation}
The sum runs over all fermion species (i.e., over the charged leptons and the
quarks of all colors and flavors), with $q_{f}$ being the charge and $b_{f}$ the
Lorentz-violating parameter for a given species. However, the radiatively induced
Chern-Simons term is proportional to exactly the quantity $\sum_{f}q_{f}^{2}
b_{f}$, so it would be no surprise if $\Delta k_{CS}$
were zero when all contributing particles were
considered.

On the other hand, no such cancelation between the species would be expected at
second
order in $b$. If the various $b_{f}$ terms do indeed arise from the expectation
value of one common field coupled to all the species of fermions, then all the
Lorentz-violating vectors should be aligned. So there are still only
two photon mass
parameters in the theory,
\begin{eqnarray}
m_{\parallel}^{2} & = & -\frac {\sum_{f}q_{f}^{2}b_{f}^{2}}{8\pi^{2}} \\
\label{eq-mperp}
m_{\perp}^{2} & = & -\frac {\sum_{f}q_{f}^{2}b_{f}^{2}}{24\pi^{2}}.
\end{eqnarray}
Even if this hypothesis for the origin of the $b$ parameters is incorrect and
these vectors are not collinear, the
minimum mass scale induced by the theory should be (barring some special
cancelation between the differing parallel and perpendicular components, in the
presence of both timelike and spacelike $b_{f}$ coefficients) roughly
the $m_{\perp}$ given by (\ref{eq-mperp}).

The best bounds on the photon mass from direct measurement come from observations
made by the Pioneer 10 probe near Jupiter~\cite{ref-davis}.
The upper limit on the mass is $6\times10^{-25}$ GeV. (Solar Probe measurements
of the sun's
magnetic field could improve this direct bound by at least an order of
magnitude~\cite{ref-kost17}.)
Treating this as a constraint on $m_{\perp}$ gives
\begin{equation}
\label{eq-blimit}
0\leq-\sum_{f}\left(\frac{q_{f}}{e}\right)^{2}b_{f}^{2} \lesssim 10^{-45}\,
({\rm GeV})^{2}.
\end{equation}
The lower limit is theoretical, and the upper limit is the experimental one. If
the $b_{f}$ all have a common origin, arising from a single $\langle A_{5}^{\mu}
\rangle$, then both these limits are completely rigorous. If not, then they are
valid again assuming that there are no special cancellations. These limits on
$b$ are comparable to or less restrictive than the best limits obtained from
direct observations of electrons, muons, and nucleons (with the limits for the
proton and neutron translating into bounds for the up and down quarks).
However, the bounds given here are much tighter than any present limits
on the $b$ coefficients for heavier particles like the tau or heavy quarks. So
the result (\ref{eq-blimit}) is best interpreted as a rough bound on the scale
of $b$-type Lorentz violation for heavy particles.

There are some much more stringent bounds on the photon mass, taken from
studies of fields on astrophysical scales. The strongest limits, which give
$m_{\gamma}\lesssim 3\times 10^{-36}$ GeV, are based on the properties of magnetic
fields on galactic scales~\cite{ref-chibisov}. Such limits involve assumptions
that may be violated in the presence of Lorentz violation. Limits based on
observations of astrophysical plasmas (e.g. the limits found in~\cite{ref-barnes},
which are conservatively at the $m_{\gamma}\lesssim 10^{-29}$ GeV level) are
likely to be more robost against the effects of Lorentz violation, although there
are still questions about these estimates. We shall not use these bounds here,
but if they were known to be valid, they would generate correspondingly stronger
limits on the $b_{f}$.

We should emphasize again that the
condition we have found on the fermionic $b_{f}$ terms is model-dependent in a
somewhat subtle way. While the physics of a classical theory is
determined entirely by the Lagrangian, the definition of a quantum theory
involves additional elements. In particular, we must give a regularization
prescription for such a theory.
In this case, the model dependence of our limit comes
directly from a dependence on how the theory is regulated. In the naive
perturbative
Feynman diagram expansion, the integral that gives rise to (\ref{eq-Pib2})
is finite and unambiguous. If no regulator is imposed, then the resulting
photon self-energy is not gauge invariant. This is the case that we have been
considering. However, it is
still possible to introduce a regulator for this integral.
Such a regulator would arise naturally if the theory were considered
nonperturbatively in $b$, because in that case, all the ${\cal O}(e^{2})$ terms in
the self-energy
would come from the same Feynman diagram; they should therefore all
be regulated in the same fashion. If a straightforward dimensional regulator is
used, then the nonzero contribution to $\Pi^{\mu\nu}_{b^{2}}$ (which arises from
a surface term) goes away, and the theory is gauge symmetric. In this case, the
role of the regulator is not to render an apparently infinite result finite, but
only to enforce gauge invariance. So if we do not insist that all the radiative
corrections be gauge invariant, then there is no reason to use a regulator for the
${\cal O}(e^{2}b^{2})$ terms, and our limits on
the $b_{f}$ apply. However, whether the theory is regulated in this fashion
is ultimately a question that can only be answered experimentally; our limits
are then most relevant if we have independent reasons to believe that
gauge symmetry may be broken by quantum corrections.

Our calculations have led to a model-dependent bound on a combination of
the $b_{f}$ coefficients for all the charged fermions in the standard model. The
bound is based on the fact that radiative corrections involving two powers of
these Lorentz-violating parameters may lead to gauge-noninvariant photon mass
terms. The bound on the $b_{f}$ then comes from comparison with
experimental constraints on the size of the photon mass. When only directly
determined bounds on the photon mass are used, we obtin limits on the $b_{f}$
at the $10^{-23}$--$10^{-22}$ GeV level; these
are much better than any other bounds presently
existing for the standard model's heavy fermions.

\section*{Acknowledgments}
The author is grateful to V. A. Kosteleck\'{y} for many helpful discussions.
This work is supported in part by funds provided by the U. S.
Department of Energy (D.O.E.) under cooperative research agreement
DE-FG02-91ER40661.

\appendix

\section*{Appendix: Waves with Lorentz-Violating Mass Gaps}

In this appendix, we derive the dispersion relations for propagating waves
when the Lagragian contains a photon mass
interaction of a particular type. A general SME
photon mass term will have the form $M^{\mu\nu}A_{\mu}A_{\nu}$. In the Proca case,
$M^{\mu\nu}=\frac{1}{2}m_{\gamma}^{2}g^{\mu\nu}$. Here, we shall consider more
general possibilities, although not the most general symmetric
mass matrix. Instead, we shall examine what occurs if there is a frame in
which the mass
matrix is diagonal, and only one of the four diagonal terms in
$M^{\mu}\,_{\nu}$ is different from the others, so that there exists a single
preferred direction in spacetime. This framework
includes all the examples mentioned in this
paper as special cases.

The equation of motion in the absence of sources is
\begin{equation}
\label{eq-EOM}
\partial^{\mu}F_{\mu\nu}+2M^{\mu}\,_{\nu}A_{\mu}=0.
\end{equation}
Taking a wave Ansatz for the field, $A\propto e^{-ik^{\mu}x_{\mu}}=
e^{i\left(\vec{k}\cdot\vec{x}-\omega t\right)}$, this becomes
\begin{equation}
\label{eq-EOMk}
\left(-\omega^{2}+\vec{k}^{2}\right)A_{\nu}+k_{\nu}\left(k^{\mu}A_{\mu}\right)+
2M^{\mu}\,_{\nu}A_{\mu}.
\end{equation}
In addition, there is a gauge-fixing condition
that comes from differentiating (\ref{eq-EOM}); since
$\partial^{\mu}\partial^{\nu}F_{\mu\nu}=0$, the
condition $M^{\mu}\,_{\nu}\partial^{\nu}A_{\nu}=0$ holds also. We may use this
equation to eliminate the $k^{\mu}A_{\mu}$ in (\ref{eq-EOMk}).

Now there are two cases to consider. In the first case, the preferred direction
is timelike.  With $M^{0}\,_{0}=\frac{1}{2}m_{0}^{2}$ and $M^{j}\,_{k}=
\frac{1}{2}\delta_{jk}m_{1}^{2}$, the gauge-fixing condition is
\begin{equation}
\label{eq-gfixingcase1}
m_{0}^{2}\omega A_{0}-m_{1}^{2}\vec{k}\cdot{\vec A}=0.
\end{equation}
Substituting
\begin{equation}
\label{eq-A0case1}
\omega A_{0}=\frac{m_{1}^{2}}{m_{0}^{2}}\vec{k}\cdot\vec{A}
\end{equation}
into (\ref{eq-EOMk}) gives the dispersion relations; these depend on whether
$\vec{k}$ and $\vec{A}$ are parallel or perpendicular:
\begin{equation}
\label{eq-dispcase1}
\omega^{2}=\left\{
\begin{array}{ll}
\vec{k}^{2}+m_{1}^{2}\,, & {\rm if}\, \vec{A}\perp\vec{k} \\
\frac{m_{1}^{2}}{m_{0}^{2}}\vec{k}^{2}+m_{1}^{2}\,, & {\rm if}\,
\vec{A}\parallel\vec{k}
\end{array}
\right..
\end{equation}
The time component of the field can then be determined from (\ref{eq-A0case1}).

The dispersions relations (\ref{eq-dispcase1}) reduce to those of a
Lorentz-invariant theory of massive vector particles if $m_{0}^{2}=m_{1}^{2}$.
However, if $m_{0}^{2}=0$, as in ${\cal L}_{m_{\gamma}}'$, the expression for the
frequency of the longitudinal component becomes singular.
Looking back at (\ref{eq-gfixingcase1}), it is clear that $\vec{k}\cdot
\vec{A}$ must actually vanish in this case, so there are only two propagating
modes if $m_{0}^{2}=0$.

The second case, in which rotation invariance is broken, is slightly more
complicated. Taking $M^{0}\,_{0}=M^{1}\,_{1}=M^{2}\,_{2}=\frac{1}{2}m_{0}^{2}$
and $M^{3}\,_{3}=\frac{1}{2}m_{3}^{2}$ (for the radiatively-induced masses
discussed in this paper, $m_{3}=\sqrt{3}m_{0}$), the gauge-fixing condition
becomes
\begin{equation}
\omega A_{0}=\vec{k}\cdot\vec{A}+\left(\frac{m_{3}^{2}}{m_{0}^{2}}-1\right)
k_{3}A_{3}.
\end{equation}
Inserting this into (\ref{eq-EOMk}) gives a matrix equation that must be
diagonalized to obtain the eigenmodes of propagation and the corresponding
frequency eigenvalues. This is straightforward to do, but the results are
rather unconventional. In fact, the dispersion relations are
\begin{equation}
\label{eq-dispcase2}
\omega^{2}=\left\{
\begin{array}{ll}
\vec{k}^{2}+m_{0}^{2}\,, & {\rm if}\,\vec{A}\perp\hat{e}_{3} \\
k_{1}^{2}+k_{2}^{2}+\frac{m_{3}^{2}}{m_{0}^{2}}k_{3}^{2}+m_{3}^{2}
\,, & {\rm if}\, \frac{\vec{A}_{\perp}}{A_{3}}=\frac{k_{3}\vec{k}_{\perp}}
{m_{0}^{2}+k_{3}^{2}}
\end{array}
\right.,
\end{equation}
where $\vec{k}_{\perp}$ and $\vec{A}_{\perp}$ are the projections of the
vectors $\vec{k}$ and $\vec{A}$
perpendicular to the $z$-direction, just as $k_{3}$ and $A_{3}$ are their
(scalar) projections along that direction.

The basis of polarization states we have found
is not orthogonal. If the mass parameters are
negligible compared to the components of $\vec{k}$, then the mode with the
unconventional dispersion relation is essentially logitudinally polarized.
However, the other modes are not necessarily transverse; their polarization
vectors are normal to $\hat{e}_{3}$, not to $\vec{k}$. In general, a transversely
polarized wave will be a superposition of two normal modes with different
frequencies.

In either of the two cases considered here, the frequency $\omega$ is guaranteed
to be real, as long as each of the diagonal terms in $M^{\mu}\,_{\nu}$ is
positive or zero. However, if any of these mass squared parameters is negative,
then there will be values of $\vec{k}$ for which $\omega$ is complex, meaning
that there are unstable runaway solutions. These solutions grow exponentially with
time. For photon masses generated radiatively from $b$, the instabilities exist if
$b^{2}>0$.

Both sets of dispersion relations, (\ref{eq-dispcase1}) and (\ref{eq-dispcase2}),
support superluminal propagation, provided the
``spacelike'' mass parameter $m_{1}$ or $m_{3}$ is greater than $m_{0}$. In that
case, the upper limit on achievable speeds in the theory is $\frac{m_{1}}{m_{0}}$
or $\frac{m_{3}}{m_{0}}$, as appropriate.
However, this maximum speed is only approached by
longitudinally polarized (or nearly longitudinally polarized) waves. If the
coupling to charged matter remains conventional, then all interactions with this
superluminal mode are suppressed by powers of the small mass parameters.
(Moreover, it is natural to theorize that the results stated in this paragraph
can be straightforwardly generalized
 to any theory with a small, symmetric, positive definite
$M^{\mu}\,_{\nu}$.) In the timelike case, the propagation speed for
longitudinally polarized waves diverges as $m_{0}^{2}\rightarrow 0$, and this
is related to the fact that the $m_{0}^{2}=0$ theory possesses an instantaneous
Coulomb interaction~\cite{ref-gabadadze,ref-dvali}.

\end{document}